\documentclass[11pt]{article}

\usepackage[margin=1in]{geometry}
\usepackage{fancyhdr}
\usepackage{graphicx} % images
\usepackage{siunitx} % units
\usepackage{booktabs} % more beautiful tables

\usepackage[
backend=biber,
maxnames=6,
bibstyle=numeric,
hyperref=true,
sorting=none,
autocite=superscript
]{biblatex}
\usepackage{csquotes}
\usepackage[toc]{glossaries} % handles nomenclature
\usepackage{glossary-longbooktabs}

\newsavebox\dummy
\newcolumntype{H}{>{\begin{lrbox}{\dummy}}c<{\end{lrbox}}@{}}

\newglossarystyle{my-breaking-long3col-booktabs}{%
  \glspatchLToutput
  \setglossarystyle{long4col-booktabs}%
    {\begin{longtable}{p{3\glspagelistwidth}p{\glsdescwidth}lH}}%
    {\end{longtable}}%
}
\setglossarystyle{my-breaking-long3col-booktabs}
\newglossary[]{symbol}{smb}{sym}{Symbols}[] % https://www.dickimaw-books.com/latex/thesis/html/makeglossaries.html#sec:newglossary

\usepackage{hyperref}

\setacronymstyle{long-short}
\newacronym{LJ}{LJ}{Lennard-Jones}
\newacronym{iLJ}{ILJ}{implicit Lennard-Jones}
\newacronym{thetaS}{$\theta$-solvent}{theta-solvent}
\newacronym{SASA}{SASA}{solvent-accessible surface area}
\newacronym{FENE}{FENE}{finitely extensible nonlinear elastic}
\newacronym{dpd}{DPD}{dissipative particle dynamics}
\newacronym{GLE}{GLE}{generalized Langevin equation}
\newacronym{SCF}{SCF}{self-consistent field}
\newacronym{PDMS}{PDMS}{polydimethylsiloxane}
\newacronym{PET}{PET}{polyethylene terephthalate}
\newacronym{MC}{MC}{Monte Carlo}
\newacronym{DPD}{DPD}{dissipative particle dynamics}
\newacronym{PBC}{PBC}{periodic boundary conditions}
\newacronym{MD}{MD}{molecular dynamics}
\newacronym{SI}{SI}{\textit{système international (d'unités)}}
\newacronym{KG}{KG}{Kremer-Grest}
\newacronym{MMT}{MMT}{Miller-Macosko Theory}
\newacronym{Euler}{Euler}{\textit{Erweiterbarer, Umweltfreundlicher, Leistungsfähiger ETH-Rechner}}
\newacronym{GPL-v3+}{GPL-v3+}{GNU General Public License v3}
\newacronym{MEHP}{MEHP}{Maximum Entropy Homogenisation Procedure}

\newglossaryentry{thetaT}{name={$\theta$-temperature},
symbol={\ensuremath{T_{\theta}}},
description={temperature at which a solvent acts as a $\theta$-solvent}}

\newglossaryentry{Ree}{name={mean square end-to-end distance},
symbol={\ensuremath{\langle R_{ee}^2 \rangle}},
description={mean squared Euclidean distance between the position of the end beads of chains (ignoring cross-linkers in case of networks)}}

\newglossaryentry{Rg}{name={mean square radius of gyration},
symbol={\ensuremath{\langle R_{g}^2 \rangle}},
description={mean squared Euclidean weighted distance of each bead to the centre of mass of the chain (ignoring cross-linkers in case of networks)}}

\newglossaryentry{N}{name={\ensuremath{N}},
symbol={\ensuremath{N}},
description={number of monomer units in a chain}}

\newglossaryentry{wsol}{name={\ensuremath{w_{\mathrm{sol}}}},symbol={\ensuremath{w_{\mathrm{sol}}}},description={soluble chain fraction}}
\newglossaryentry{wdang}{name={\ensuremath{w_{\mathrm{dang}}}},symbol={\ensuremath{w_{\mathrm{dang}}}},description={dangling chain fraction}}

\newglossaryentry{phiD}{name={\ensuremath{\Phi_D}},symbol={\ensuremath{\Phi_D}},description={mass fraction of dangling chains}}
\newglossaryentry{phiB}{name={\ensuremath{\Phi_{el}}},symbol={\ensuremath{\Phi_{el}}},description={mass fraction of network backbone}}

\newglossaryentry{p}{name={\ensuremath{p}},symbol={\ensuremath{p}},description={extent of reaction}}

\newglossaryentry{pgel}{name={\ensuremath{p_{\mathrm{gel}}}},symbol={\ensuremath{p_{\mathrm{gel}}}},description={gelation point}}

\newglossaryentry{r}{name={\ensuremath{r}},symbol={\ensuremath{r}},description={stoichiometric imbalance}}

\newglossaryentry{f}{name={\ensuremath{f}},symbol={\ensuremath{f}},description={cross-link functionality}}

\newglossaryentry{G}{name={equilibrium shear modulus},
symbol={\ensuremath{G_{\text{eq}}}},
description={stress relaxation modulus for a viscoelastic solid \cite{rubinstein_polymer_2003}}}

\newglossaryentry{D}{name={polydispersity},symbol={\DJ},description={relation of weight-average and number-average molar mass}}

\newglossaryentry{taurlx}{name={relaxation time},
symbol={\ensuremath{\tau_{\text{rlx}}}},
description={characteristic relaxation time}}

\newglossaryentry{kb}{name={Boltzmann constant},
symbol={\ensuremath{k_B}},
description={thermodynamic constant, equal to \SI{1.380649e-23}{\joule\per\kelvin}\cite{noauthor_codata_nodate}}}

\newglossaryentry{phi}{name={dilution},
symbol={\ensuremath{\phi}},
description={measure of solvent fraction}}

\newglossaryentry{stress}{name={stress},
symbol={\ensuremath{\sigma}},
description={measure of the internal forces that neighbouring particles of a continuous material exert on each other},
plural={stresses}}

\newglossaryentry{strain}{name={strain},
symbol={\ensuremath{\gamma}},
description={measure of deformation}}

\newglossaryentry{T}{name={temperature},
symbol={\ensuremath{T}},
description={temperature,
measure of kinetic energy}}

\newacronym{LAMMPS}{LAMMPS}{Large-scale Atomic/Molecular Massively Parallel Simulator\cite{plimpton_fast_1995}}
\newacronym{Gromacs}{GROMACS}{GROningen MAchine for Chemical Simulations\cite{abraham_gromacs_2025}}
\newacronym{COGNAC}{COGNAC}{COarse-Grained molecular dynamics program by NAgoya Cooperation\cite{noauthor_cognac_nodate}}
\newacronym{OCTA}{OCTA}{Open, flexible and expandable system OCTA\cite{noauthor_octa_nodate}}
\newacronym{CP2K}{CP2K}{CP (Car-Parrinello = ab initio MD) code for the new Millenium; note that CP2K\cite{kuhne_cp2k_2020} didn't implement the CP method}
\newacronym{PSP}{PSP}{Polymer Structure Predictor\cite{noauthor_ramprasad-grouppsp_2025}}

\makeglossaries

\begin{document}

\title{pylimer-tools: A Python Package for Generating and Analyzing Bead-Spring Polymer Networks}

\author{Tim Bernhard\textsuperscript{1,2}, Fabian Schwarz\textsuperscript{3}, Andrei A. Gusev\textsuperscript{1,2}}

\date{}

\maketitle

\begin{abstract}
The Python package pylimer-tools is a comprehensive toolkit for computational studies of polymer networks, particularly bead-spring networks.
The package provides functionality to generate polymer networks using Monte Carlo (MC) procedures and analyze their structural and mechanical properties.
Key features include detection of loops, reduction of the network to its ground state energy both with and without entanglements by the Force Balance procedure, and thereafter computing the soluble and dangling fractions of network strands, as well as the equilibrium shear modulus.
The toolkit supports analysis of structures generated both internally and by external simulation software such as LAMMPS.
The package implements theoretical frameworks including Miller-Macosko theory and provides a dissipative particle dynamics (DPD) simulator with slip-spring entanglement modeling.
Built with C++ for performance and exposed through Python bindings, pylimer-tools addresses the need for specialized tools in computational polymer science.
\end{abstract}


\noindent\textbf{Affiliations:}\\
\textsuperscript{1} Laboratory for Nanometallurgy, D-MATL, ETH Zürich\\
\textsuperscript{2} D-MATL, ETH Zürich\\
\textsuperscript{3} Structural Chemistry, Department of Chemistry -- Ångström, Uppsala University

\section{Introduction}

The computational study of polymeric systems has become increasingly important, driven by advances in computational hardware that enable simulation of increasingly large and architecturally complex polymeric systems.
Such simulations can reproduce experimental data and show promise for the architectural design of polymer networks, potentially accelerating research and development in the polymer industry.

When simulating a polymer system, a typical workflow involves:
\begin{enumerate}
  \item setting up an initial structure through a \gls{MC} procedure\cite{gusev_numerical_2019,gusev_molecular_2019}, a hierarchical strategy\cite{komarov_highly_2007,zhang_equilibration_2014,}, or a \gls{MD} diffusion-collision-based generator\cite{svaneborg_connectivity_2008,gusev_molecular_2022,gula_computational_2020},
  \item running simulations using \gls{MC}, \gls{DPD}, or \gls{MD} methods,
  \item analyzing and visualizing the results.
\end{enumerate}

The \gls{KG} bead-spring model\cite{grest_molecular_1986,grest_vectorized_1989,kremer_dynamics_1990} provides an optimal balance between structural detail and computational efficiency for studying the mechanical properties of polymers.
The computer codes for \gls{KG} simulations are usually highly optimized to support parallel computations on both CPUs and GPUs in order to maximize the productivity of the researchers while allowing for the study of larger systems and longer trajectories, as is required in particular for polymeric systems.

%To our current knowledge, there are no simulation programs specifically designed for polymeric systems.
%Instead
Usually, general \gls{MD} simulators or programs for atomistic simulations are used for such simulations of polymeric systems.
Examples include \gls{LAMMPS}, \gls{Gromacs}, \gls{COGNAC} (part of the \gls{OCTA} project), or \gls{CP2K}.
As a consequence of the programs being general, even though these programs provide many possibilities to output various properties of the simulated system, they are not specific for polymeric systems.
Therefore, various network properties such as \gls{p} (\glsdesc{p}), \gls{pgel} (\glsdesc{pgel}) or \gls{r} (\glsdesc{r}) have to be computed externally, through clever use of the commands provided by the program, or by providing a custom addition to the program.
This becomes even more pronounced for analyzing the network structure, such as determining \gls{phiD} (\glsdesc{phiD}), \gls{phiB} (\glsdesc{phiB}), or the distribution and length of loops in the structure, as these are properties that are key only for polymeric simulations and therefore have no immediate applications in other \gls{MD} projects, such as simulating metals, crystals, glasses, or galaxies.

Some specialized tools exist for specific aspects of polymer simulation.
For example, the Pizza.Py toolkit offers a Python\cite{van_rossum_python_2009} programming interface for \gls{LAMMPS}, which can be used to create and manipulate structures.
The MDAnalysis Python package\cite{michaud-agrawal_mdanalysis_2011,gowers_mdanalysis_2016} provides also provides a toolkit to read and process \gls{MD} data from various simulation programs.
For generating worm-like three-dimensional chains, the PolymerCpp Python package\cite{stefko_polymercpp_2020} is available.
A hierarchy of polymer models can be generated using the \gls{PSP}.
And the toolbox by \citeauthor{barrett_tj-barrettpolymer-toolbox_2024}\cite{barrett_tj-barrettpolymer-toolbox_2024} provides polymer generation functionality based on molecular templates, and methods to compute the static structure factor and the radial distribution function.

% For some of the functionality named above, such as determining the \gls{phiD}, we are not aware of any available codebase that provides it.
% Because of the absence of a comprehensive public toolkit encompassing all the required polymer properties, and because of the challenges associated with the interoperability of the aforementioned packages, we have undertaken the implementation of these functionalities ourselves as necessary for our prior publications.
% Now, we are making our code open source and introduce in this article the Python package that has been developed for these calculations and for an easier integration with preparation of plots and figures. 

However, these tools often lack interoperability and comprehensive coverage of polymer-specific analysis methods.
To address these limitations, we have developed \texttt{pylimer\_tools}, a comprehensive Python package that consolidates polymer-specific computational tools into a unified, efficient framework designed specifically for bead-spring polymer network studies.

\section{Implementation and architecture}

This package, \texttt{pylimer\_tools}, is implemented primarily in C++ for optimal performance of computationally intensive calculations and input/output operations.
Python bindings are provided through PyBind11\cite{noauthor_intro_nodate}, ensuring seamless integration with Python workflows while maintaining computational efficiency.
The package is cross-platform compatible, tested on Linux, macOS, and Windows with Python 3.8 and later versions.
Comprehensive unit tests and continuous integration ensure code quality and correctness across different platforms and use cases.

\paragraph{Core Architecture}
The software leverages the igraph library\cite{csardi_igraph_2006} to represent molecular structures as graphs, where atoms are vertices and bonds are edges.
This graph-based representation enables efficient analysis of network topology, including clustering, connectivity traversal, angle and dihedral detection, and manipulation of periodic boundary conditions.
The graph structure facilitates rapid determination of atom coordination numbers and the decomposition of systems into disconnected networks or individual polymer chains.

\paragraph{Data Structures}
The central \texttt{Universe} class encapsulates complete polymer network structures.
Users can build these programmatically by adding atoms and bonds, or import existing structures from \gls{LAMMPS} data and dump files.
A trajectory from a dump file or from a sequence of data files is represented by the \texttt{UniverseSequence} class, which allows memory-efficient on-demand reading of the respective \texttt{Universe} at a specific time-step.
The \texttt{Universe} can be decomposed into \texttt{Molecule}s, which represent the strands and chains in the structure.
After manipulating the structure, this library also provides the functionality to output the structure back into a data file compatible with \gls{LAMMPS}.
The package provides additional comprehensive file I/O capabilities, including reading various simulation output files of \gls{LAMMPS} (correlation data, vector outputs, thermodynamic logs).

\paragraph{Network Generation}
The \texttt{MCUniverseGenerator} class implements flexible Monte Carlo network generation algorithms\cite{gusev_numerical_2019,tsimouri_monte_2021} supporting end-linked and vulcanized networks, as well as complex architectures like bottle-brush or comb-like polymers\cite{tsimouri_lightweight_2024}.
The beads between the cross-links are placed by default using a Brownian Bridge process\cite{mansuy_aspects_2008,ibe_9_2013,chow_brownian_2009,martinez_thermodynamics_1996,wang_exact_2020}, though additional \gls{MC} steps can be taken to progress with the equilibration of the Gaussian springs between the beads.

\paragraph{Force Balance and Energy Minimization}
A key feature is the implementation of the \gls{MEHP}\cite{gusev_molecular_2019,gusev_numerical_2019} and the related Force Balance procedure.
There are three different implementations available.
In the class \texttt{MEHPForceRelaxation}, phantom network shear moduli can be determined efficiently with alternative, nonlinear spring potentials.
In the class \texttt{MEHPForceBalance}, the package extends the phantom network with slip-links to model entanglements, providing estimates of real network moduli.
The third implementation is given in \texttt{MEHPForceBalance2} through the Force Balance procedure with fixed entanglement links, that is, entanglements modeled as tetrafunctional cross-links, as published recently.\cite{bernhard_phantom_2025}
This is our fastest implementation of the \gls{MEHP}, but is consequently less flexible than the other two implementations.
Apart from making accurate predictions of the modulus of real networks, it also enables a systematic investigation of different entanglement detection methods and their effects on mechanical properties, for example.

\paragraph{Simulation Capabilities}
An integrated \gls{DPD} simulator with slip-spring entanglement modeling\cite{langeloth_recovering_2013,schneider_entanglements_2022} supports multi-core parallel execution via OpenMP.
Normal mode analysis for dynamic mechanical properties (stress autocorrelation, loss and storage moduli\cite{gusev_molecular_2024}) is implemented using LAPACK\cite{anderson_lapack_1999} for efficient eigenvalue computations\cite{gusev_molecular_2024}.

\paragraph{Advanced Algorithm Implementations}
An efficient loop detection algorithm based on \citeauthor{johnson_finding_1975}\cite{johnson_finding_1975} was implemented and can be used to detect loops of arbitrary order in the network.
The package includes a neighbor list implementation using spatial partitioning that optimizes performance for entanglement sampling and force calculations.
% This neighbor list is also exposed to Python and is implemented using a cell-based spatial partitioning approach, where the simulation box is divided into a grid of buckets.
% Particles are assigned to buckets based on their coordinates, and neighboring particles are efficiently identified by examining adjacent buckets within a specified cutoff distance.
%This method, commonly known as the cell list or linked-cell list algorithm, significantly reduces the computational complexity of identifying nearby particles compared to a brute-force approach

\begin{figure}[htb]
  \centering
  \includegraphics[width=0.95\linewidth]{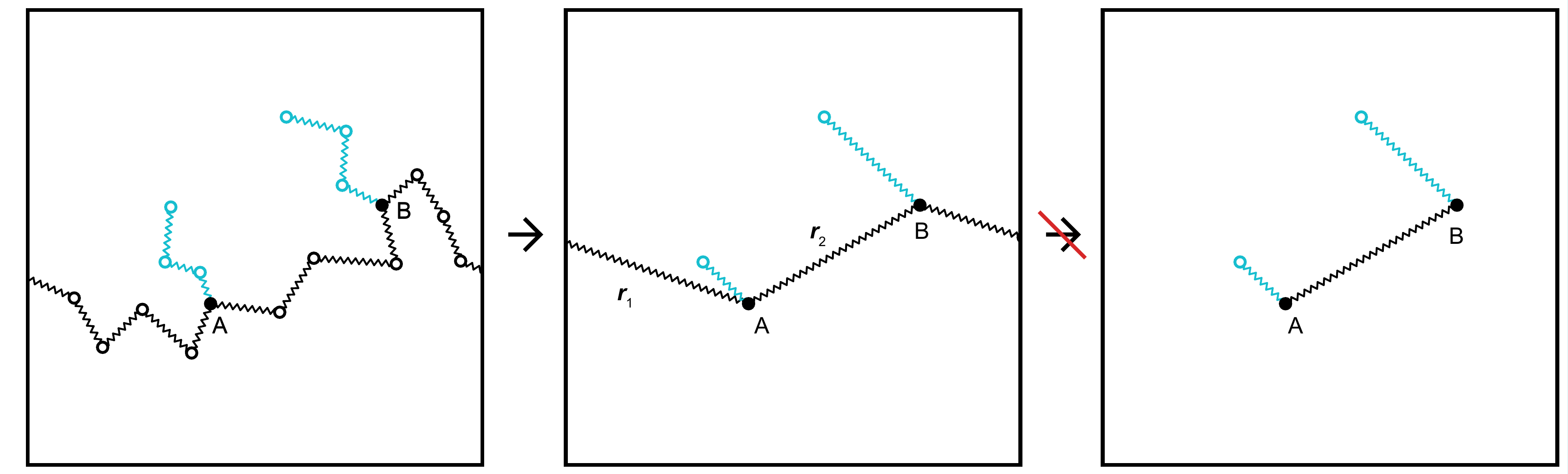}
  \caption{An illustration of the dangers of the common periodic boundary conditions for optimization procedures implemented in pylimer\_tools: when strands are converted into single springs, if the strands are longer than half the box size, the springs might collapse (step 3) even if they should not have. To prevent this, the initial state is used to derive multiples of the box to be used as offsets when computing the spring lengths. In this example, the vector $\mathbf{r}_1$ from A to B would have an offset of negative one box length.}
  \label{fig:pbc-relevance}
\end{figure}

\paragraph{Periodic Boundary Conditions}
All the simulation methods mentioned so far allow switching between two modes of \gls{PBC}.
The classical mode implements the periodic boundaries simply by using the shortest distance, as is also implemented, for example, in \gls{LAMMPS}.
However, there are some scenarios where this is not sufficient.
For example in the \gls{MEHP}, to reduce the number of degrees of freedom, multiple bonds are combined into one long spring (bifunctional beads are not represented in the degrees of freedom).
Another example is the \gls{DPD} simulations, which may lead to extraordinarily long bonds due to random fluctuations and the soft bond potential.
For these scenarios, one has to prevent atoms or beads from crossing into another periodic image (see Figure \ref{fig:pbc-relevance}).
This is achieved by deriving a periodic offset at the beginning of the procedure.
At that point, the springs have not yet replaced the bonds, and the bonds are assumed to still be shorter than half the box length.
Consequently, one can follow the graph structure to find the bonds that must have an offset to account for the periodic boundary condition.
Throughout the remainder of the procedure, this offset will remain associated with the corresponding bonds or springs. 
Apart from maintaining this constant offset, no additional steps are necessary to achieve a functional implementation of periodic boundary conditions.

\paragraph{Unit Management}
Integration with the pint library\cite{noauthor_pint_nodate} provides automatic unit conversion between LAMMPS unit systems\cite{noauthor_units_nodate} and \gls{SI} units, using conversion factors from \citeauthor{everaers_kremergrest_2020}\cite{everaers_kremergrest_2020}, preventing unit-related errors in calculations.

\paragraph{Theoretical Framework Implementation:}
Readily available methods are provided to use the \gls{MMT}\cite{miller_new_1976,macosko_new_1976} for relating the quantities \gls{p}, \gls{r}, \gls{G} \glssymbol{G} and \gls{f} of end-linked polymer networks, which were shown to be highly accurate\cite{gusev_molecular_2022}.
We have implemented the methods for predicting the \gls{wsol}, \gls{wdang}, \gls{pgel}, and both the phantom and the entangled \gls{G}.
For all of them, we go further than the previously published literature: for $\glssymbol{f}>4$, we solve the polynomial equation to predict the probability numerically.
We include the possibility of having a fraction of monofunctional chains for arbitrary \gls{f} to predict any of the relevant properties.
These derivations were validated using those published previously for the determination of \gls{wsol} for $\glssymbol{f}=3$\cite{urayama_damping_2004} and $\glssymbol{f}=4$\cite{patel_elastic_1992}.

\section{Quality control}

Software quality is ensured through a comprehensive testing framework using Catch2 for C++ components and pytest for Python code.
The test suite includes unit tests, integration tests, and performance benchmarks. %, achieving over 95\% code coverage.
Continuous integration through GitHub Actions automatically runs a representative set of tests on every code change, ensuring that new contributions do not introduce regressions or break existing functionality.
The complete test suite is run on multiple platforms (Linux, macOS, Windows) and Python versions (3.9-3.12) released before a new version of \texttt{pylimer-tools} is published on PyPI.

The testing pipeline includes the following.

\begin{itemize}
  \item \textbf{Unit tests}: Individual function and class testing with edge case validation
  \item \textbf{Integration tests}: End-to-end workflow testing including file I/O operations
  \item \textbf{Numerical validation}: Comparison with analytical solutions and published results
  \item \textbf{Performance benchmarks}: Monitoring computational efficiency for critical algorithms
  \item \textbf{Memory safety}: Leak detection and bounds check using sanitizers
\end{itemize}

Version releases follow semantic versioning with automated deployment to PyPI after passing all tests.
The package includes extensive documentation with API references, tutorials, and example workflows.
Code quality is maintained through automated formatting (clang-format, ruff) and static analysis.

For end users, the software provides clear error messages and warnings for invalid inputs or edge cases.
Example datasets and benchmark problems are included to verify the correct installation and functionality.
The test suite can be run locally using the build scripts provided, enabling users to validate their installation environment.

\section{Availability}

\subsection{Operating system}

This package is at least compatible with any MacOS, Windows and Linux version supporting Python 3.9 and newer.

\subsection{Programming language}

The core of the package is implemented in C++20 and exposed to Python 3.9+ through PyBind11 bindings.

\subsection{Additional system requirements}

The package installation requires approximately \SI{10}{\mega\byte} of disk space.
Memory and CPU requirements scale with system size: typical polymer networks with around \num{5e5} atoms require \SIrange{1}{8}{\giga\byte} RAM for a workflow that includes network generation using the \gls{MC} procedure, entanglement sampling and the Force Balance procedure to determine the \gls{G}.
Multi-core processors are recommended for DPD simulations and normal mode analysis.

\subsection{Dependencies}

Python packages that are required for basic functionality:
\begin{enumerate}
  \item \texttt{pandas}\cite{mckinney_data_2010,team_pandas-devpandas_2020}: reading simulation output files will return Pandas data frames for comfortable data handling
  \item \texttt{numpy}\cite{harris_array_2020}: for efficient vectorized calculations
  \item \texttt{pint}\cite{noauthor_pint_nodate}: for unit conversions
\end{enumerate}

Build dependencies (required only for source compilation):
\begin{enumerate}
  \item \texttt{CMake}: Cross-platform build system
  \item C++20-compatible compiler
  \item Python development headers (python3-dev on Linux)
\end{enumerate}

External libraries (automatically downloaded during build if not found on the system):
\begin{enumerate}
  \item \texttt{igraph}\cite{csardi_igraph_2006}: Graph algorithms and data structures
  \item \texttt{Eigen}\cite{guennebaud_eigen_2010}: Linear algebra operations
  \item \texttt{nlopt}\cite{johnson_stevengjnlopt_2022}: Nonlinear optimization algorithms
  \item \texttt{pybind11}\cite{noauthor_intro_nodate}: Python-C++ binding framework
  \item \texttt{Spectra}\cite{qiu_spectra_2015}: Eigenvalue computation for sparse matrices
\end{enumerate}

\subsection{List of contributors}

\begin{enumerate}
  \item Tim Bernhard, ETH Zürich: main developer. These codes are the means used for all results published for and during his PhD
  \item Andrei A. Gusev, ETH Zürich: initial Mathematica and C versions of the \gls{MC} network generator and the \gls{MEHP} force relaxation codes that had a large influence on the current C++ codes
  \item Fabian Schwarz, Uppsala University: initial Matlab versions of the \gls{MC} network generator and the \gls{MEHP} force relaxation codes that had a large influence on the current C++ codes
  \item Jorge Ramirez, Universidad Politécnica de Madrid: multiple-tau autocorrelation\cite{ramirez_efficient_2010} C codes used as a basis of the C++ implementation
  \item Martin Kröger, ETH Zürich: discussions regarding implementations of periodic boundary conditions for small systems
\end{enumerate}

\subsection{Software location}

\paragraph{Archive}

\begin{itemize}
  \item \textbf{Name:} Zenodo
  \item \textbf{Persistent identifier:} \url{https://doi.org/10.5281/zenodo.16633352}
  \item \textbf{Licence:} Creative Commons Attribution 4.0 International
  \item \textbf{Publisher:} Tim Bernhard
  \item \textbf{Version published:} 0.3.1
  \item \textbf{Date published:} 31/07/2025
\end{itemize}

\paragraph{Code repository}

\begin{itemize}
  \item \textbf{Name:} GitHub
  \item \textbf{Persistent identifier:} \url{https://github.com/GenieTim/pylimer-tools}
  \item \textbf{Licence:} GPL-3.0 license
  \item \textbf{Date published:} 31/07/2025
\end{itemize}

\subsection{Language}

English

\section{Reuse potential}

Other researchers can use this package already if their sole intent is to read \gls{LAMMPS} output files, since it provides fast and ergonomic functions to do so, or if they use the \gls{MMT} thanks to the respective formulas being implemented and easily available.
The real benefit becomes apparent for polymer researchers who do \gls{KG} \gls{MD} simulations or other studies using bead-spring polymer models.
For these researchers, this package provides a large number of tools that can be used, from generating structures using \gls{MC} procedures, writing these files for usage with \gls{LAMMPS}, using the structures with the built-in \gls{DPD} simulator or analyzing the loop distributions, end-to-end distances, bond lengths, or reducing it to its minimum energy state using the Force Balance procedure to compute the \gls{G}.

As the software is licensed under an open source license, further adaptations are also possible.
Feature requests and problems or issues can be submitted by raising an issue in the GitHub repository. The distribution includes a detailed manual that covers the installation and use of the software.

\section{Acknowledgements}

The authors thank Martin Kröger for the discussion on alternative implementations of \gls{PBC}.% and Fabian Schwarz for providing original Matlab scripts.
TB also wishes to thank Szabolcs Horvát for his positive interactions on the use and extension of \texttt{igraph}, and Jorge Ramirez for providing access to his implementation of the multiple-tau autocorrelation and answering questions about its license.

\section{Funding statement}

The authors gratefully acknowledge financial support from the
Swiss National Science Foundation (SNSF project 200021\_204196).

\section{Competing interests}

The authors declare that they have no competing interests.

\section{References}

\printbibliography[heading=none]

\printglossary[title={Notation}]

\end{document}